# Colorization of Optically Transparent Surfactants to Track Their Movement in Biphasic Systems Used for Differentiation of Nanomaterials

Błażej Podleśny[a,*], Łukasz Czapura[a], Dawid Janas[a,*]

[a] *Department of Organic Chemistry, Bioorganic Chemistry and Biotechnology, Silesian University of Technology, B. Krzywoustego 4, 44–100 Gliwice, Poland*

[*] Corresponding authors: Blazej.Podlesny@polsl.pl (B.P), Dawid.Janas@polsl.pl (D.J.)

**Abstract**

Aqueous two-phase extraction (ATPE) is a versatile method for the purification of numerous chemical compounds and materials, ranging from proteins and nucleic acids to cell organelles and various nanostructures. However, despite its widespread use, the underlying extraction mechanism remains unclear, which significantly reduces the utility of ATPE. Many types of surfactants are often added to biphasic systems to enhance the extraction of analytes between phases. Although their role in this process is crucial, it is not entirely understood. In this work, to fill this gap, we adapt and refine a nearly two-hundred-year-old chemical technique for the detection of bile salts in urine, referred to as Pettenkofer's test and monitor the partitioning of single-walled carbon nanotubes (SWCNTs) by ATPE. This approach enabled us to tint the otherwise transparent bile salt surfactants to precisely track their distribution and concentration in the biphasic system, thereby unravelling the modus operandi of this popular purification technique.

## Introduction

Single-walled carbon nanotubes (SWCNTs) are carbon allotropes that have consistently attracted scientific interest since their discovery. Although all SWCNTs have a cylindrical shape and are composed of a hexagonal carbon sheet, the spatial arrangement of atoms can vary significantly[1]. Consequently, each SWCNT type exhibits different properties, posing a major challenge for their implementation, as there is no commercially viable method for synthesizing monochiral SWCNT products in sufficiently large amounts[2]. To overcome this problem, various post-synthesis separation techniques have been developed to isolate SWCNTs with specific structure and hence properties. Among these, water-based methods such as density gradient ultracentrifugation[3], electrophoresis[4], chromatography[5], and aqueous two-phase extraction (ATPE)[6] are widely employed. The latter, which relies on the partitioning of SWCNTs between two immiscible aqueous solutions, is particularly effective for this purpose as it can deliver fractions enriched with SWCNTs of specific conductivity type[7], chirality[8], and handedness[9].

However, due to the highly hydrophobic nature of SWCNTs, their use in a liquid medium requires the application of dispersing agents. Bile salt surfactants, particularly sodium cholate (SC) and sodium deoxycholate (SDC), are often used to achieve individualized, stable SWCNT suspensions[2]. In ATPE, such suspensions are introduced into a biphasic system, typically composed of dextran (DEX) and poly(ethylene glycol) (PEG)[6]. Depending on the type of surfactant coating, SWCNTs preferentially partition into one of the two phases. For example, with SC or SDC, SWCNTs typically migrate into the DEX-rich bottom phase[10]. To achieve SWCNT separation, surfactants increasing the SWCNT affinity to the top phase (e.g., sodium dodecyl sulfate (SDS) or Triton X-100)[10–12] are gradually introduced. As a result, SWCNTs can be collected stepwise from the top phase. Analogously, SWCNT dispersions encapsulated with surfactants that prefer the top phase can be prepared, and then the SWCNTs are harvested step by step in the bottom phase.

Despite the conceptual simplicity of this approach, the mechanism governing the separation of SWCNTs remains unresolved. One particularly essential but largely overlooked aspect is the distribution of surfactants between the two phases. For an unknown reason, the addition of surfactants to the biphasic system causes the migration of SWCNTs between the phases. As previously stated, bile salt surfactants promote the relocation of SWCNTs to the bottom phase. Hence, is the concentration of these surfactants higher therein, and this increased preference for the bottom phase is the reason why they facilitate the downward extraction of SWCNTs? Considering the dual function of surfactants in this approach (playing the roles of SWCNT solubilizers and partitioning modulators), it is desirable to answer such questions and pinpoint their role in order to come closer to elucidating the workings of ATPE to improve its utility for purification of a broad spectrum of materials, which it can process[13,14].

Unfortunately, this is a challenging endeavor. Although the ATPE samples contain a limited number of components (usually two phase-forming polymers, up to three surfactants, SWCNTs, and water), no straightforward analytical method exists for the quantitative determination of surfactants. These compounds are typically colorless and exhibit overlapping broad absorption bands with other ATPE components. Therefore, analysis thereof using absorption spectroscopy is not viable. Concomitantly, nuclear magnetic resonance and mass spectroscopy are also not well-suited for quantitative analysis of surfactants present in the matrices of phase-forming components (PEG and Dextran), which are polydisperse and made of the same elements as regular surfactants used in the ATPE framework. Taking into account the widespread use of surfactants, particularly in the context of nanomaterial solubilization and modification[2,15,16], it would be beneficial to establish a platform for their selective and straightforward determination to improve the comprehension of the phenomena at the nanoscale.

In this work, we present a colorimetric method for the quantitative determination of surfactants in complex chemical environments. By adapting and improving the relatively unpopular

Pettenkofer test, first published in 1844[17,18], we elucidated how to quantify the abundance of bile salt surfactants, such as sodium cholate, directly in the biphasic system composed of poly(ethylene glycol), dextran, SWCNTs, water, etc. The developed method proved to be very robust and compatible with various chemicals present in water. Moreover, it facilitates the detection of bile salt surfactants with the naked eye. Capitalizing on this achievement, we successfully deciphered the mechanism of separating SWCNTs via ATPE with bile salt surfactants. The obtained results reveal meaningful surfactant-SWCNT interactions, paving the way for the development of more effective purification strategies for materials and compounds that can be processed by surfactant-assisted extraction, especially in biphasic systems.

**Results**

During original Pettenkofer's procedure, sucrose undergoes hydrolysis followed by dehydration to form furfural (FUR), (5-hydroxymethyl)furfural, or other analogous compounds[19] which then react with cholic acid to produce a colored compound[20]. Dextran used in ATPE separation is a glucose-based polymer. Hence, in an acidic environment, it also gradually undergoes hydrolysis and oxidation producing furfural and other aromatic aldehydes[21,22], which may take part in the Pettenkofer reaction, precluding its use for SC detection. The measurement of SC concentration *in situ*, either in the top or bottom ATPE phase to unravel its mechanism, using the classical Pettenkofer approach, both of which contain a certain amount of DEX[14], would not give reliable results. To overcome this problem and liberate the SC detection from the dependence on the complex polysaccharide decomposition pathway, we intentionally added aldehydes into the detection platform instead of sucrose, suspecting that the SC may react directly with them. A selection of derivatives of the simplest aromatic aldehyde, benzaldehyde, were examined, while FUR was evaluated as a reference (**Figure 1a**).

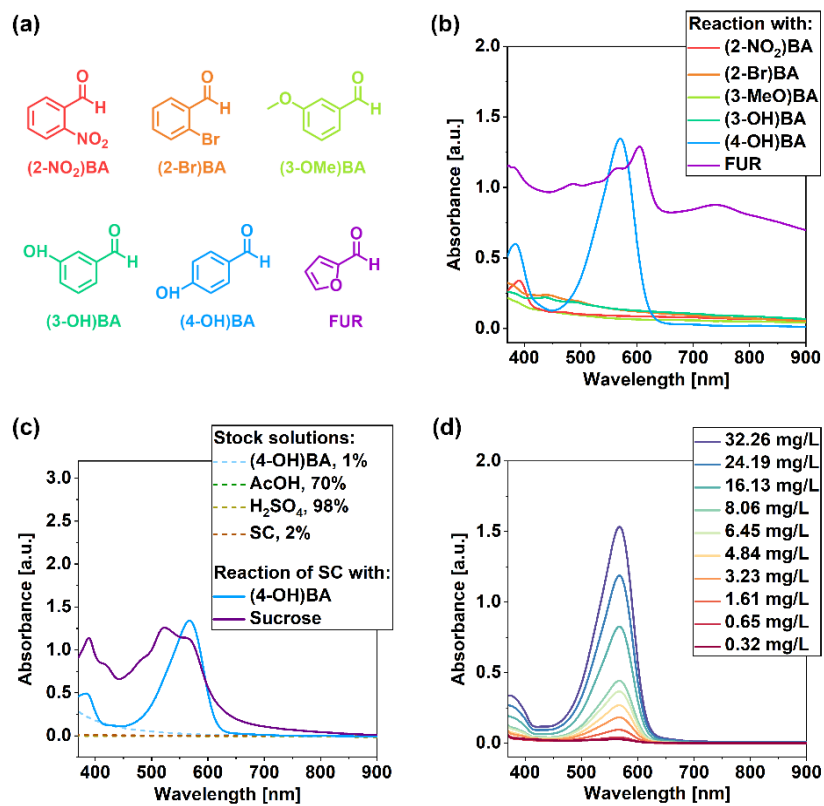

**Figure 1** (a) A selection of aromatic aldehydes chosen for conjugation with SC according to the modified Pettenkofer's test procedure, optical absorption spectra of (b) the reaction products between SC and the indicated compounds, (c) stock solutions used for the modified Pettenkofer's test and the products of reaction between SC and (4-OH)BA or sucrose, (d) the products of reaction between (4-OH)BA and SC at various indicated concentrations.

Interestingly, among the tested compounds, (4-OH)BA displayed a very promising performance as its reaction with SC in acidic environment produced a distinct optical feature positioned at ca. 570 nm (**Figure 1b**). At the same time, other benzaldehyde derivatives containing electron-donating and electron-withdrawing substituents in different positions did not afford colored products. Across the evaluated benzaldehyde analogs, (4-OH)BA was the only one possessing an electron-rich hydroxy group attached in the para position, which is considered as ring-activating[23]. Other electron-abundant compounds (3-OMe)BA and (3-OH)BA had the functional groups positioned in the meta position, which cannot increase the electron density in the aromatic system analogously[24]. Concomitantly, the (2-NO$_2$)BA and (2-Br)BA compounds were relatively electron-deficient and functionalized in close proximity to the formyl group, producing steric hindrance. It is important to note that the colored product made from (4-OH)BA and SC had a different wavelength of optical absorption maximum from the products of corresponding reactions of FUR (possible sugar decomposition product, **Figure 1b**) or sucrose (**Figure 1c**) with SC. Since FUR or sucrose (and, by extension, glucose) can be generated from DEX, the application of (4-OH)BA alleviates the problem of the concurrent side reaction of SC with DEX, making SC quantification robust. The possibility of precise determination of SC concentration is also supported by the fact that the reagents used for the improved variant of the Pettenkofer reaction do not produce any colored products in the explored spectral range **(Figure 1c)**. Last but not least, the colored product created in the reaction between (4-OH)BA and SC manifests as a relatively intensive peak in the optical absorption spectra, the intensity of which is linearly dependent on SC concentration (**Figure 1d, Figure S3**), making such quantification straightforward.

Capitalizing on the fact that (4-OH)BA turned out to be very effective for analysis of synthetic solutions containing only SC and water by making a purple colored solution (**Figure 2a**), we extended this concept to investigate much more complex environments containing SC. In the ATPE system, commonly made of immiscible PEG and DEX aqueous solutions, SWCNTs can be extracted stepwise to the bottom phase with increasing SC concentration (**Figure 2b**). In such case, SWCNTs with the smallest diameters migrate first, while larger SWCNTs require higher concentrations of SC. In the displayed example, once the volume of SC (10%, aq.) was increased from 0 μL to 450 μL in the system, the smallest available SWCNTs, i.e., (6,5) SWCNTs (d = 0.757 nm) and (8,3) SWCNTs (d = 0.782 nm), were extracted to the bottom phase **(Figure 2c)**. Further increase in the SC volume to 600 μL caused the emergence of the slightly larger (7,5) SWCNTs (d = 0.829 nm) and (8,4) SWCNTs (d = 0.840 nm) in the bottom phase. Currenttly it is unclear how surfactants promote this move in the ATPE framework.

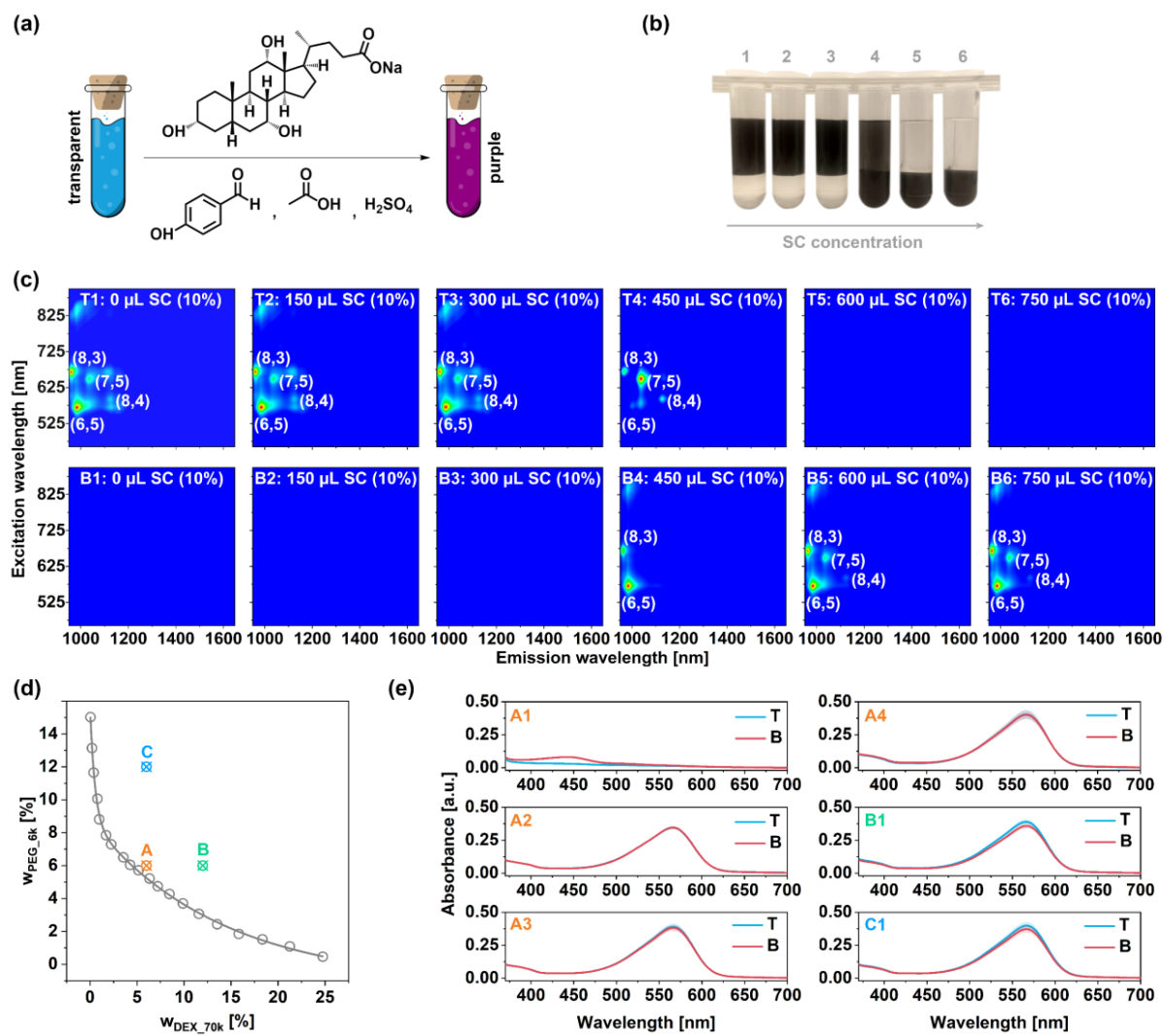

**Figure 2** (a) Generation of a colored product from SC enabling its optical detection, (b) the changed distribution of SWCNTs in the ATPE system as a function of SC surfactant concentration, (c) photoluminescence excitation-emission maps of the top and bottom phases of the corresponding samples shown in panel (b), (d) binodal curve of the biphasic system made from PEG (6 kDa) and DEX (70 kDa) with the experimental conditions explored indicated by letters (the area above the curve indicates the biphasic region conditions), (d) optical absorption spectra of the products of reactions between the top and bottom phases collected from ATPE experiments performed according to the conditions reported in **Table S3**. The solid line indicates the measured data, whereas the shaded area corresponds to experimental uncertainty determined in **Table S4**.

We analyzed three variants of DEX-PEG systems (**Figure 2d**), which are the most used in SWCNTs separation with the ATPE. **Table S3** lists the volumes of compounds combined to form these systems, which were later examined using the disclosed SC detection approach. The first series (A) corresponds to samples wherein the concentration of both phase-forming components (PEG, DEX) was the same, i.e., 6%. To investigate whether the relative amount of one phase to the other impacts the SC distribution in the biphasic systems, two other series were made: (B) with the concentrations of PEG and DEX equal to 6% and 12%, respectively, while for the (C) series these amounts were inverted giving the concentrations of PEG and DEX equal to 12% and 6%, respectively.

The samples of the top (PEG) and bottom (DEX) phases subjected to the modified Pettenkofer's test (**Figure 2d** – A1) showed only a slight increase in absorbance in case of the latter corresponding to hydrolysis and oxidation of DEX. However, this peak with minor intensity was positioned at ca. 440 nm, so it did not coincide with the spectral feature resulting from the combination of (4-OH)BA and SC. Once SC was added into the PEG-DEX system, the expected absorption band emerged at ca. 570 nm in the case of both phases (**Figure 2d** – A2) after the reaction and its intensity increased upon adding a higher amount of SC (**Figure 2d** – A3). The situation was analogous when 225 μL of SWCNT dispersion in SC aqueous solution was added (Sample A4) instead of pure 225 μL of SC aqueous solution (Sample A2). Two important conclusions can be drawn based on these outcomes. Firstly, the devised approach can be successfully used to track the distribution of SC between the phases, regardless of the complexity of the system and even in the presence of SWCNTs. The high optical density of the product formed by SC and (4-OH)BA reaction covers all signals coming from SWCNTs the absorption of which is much smaller taking into account their concentration. Secondly, it was somewhat surprising that the SC promoting the migration of SWCNTs to the bottom phases[10,11,25] turned out to have the same concentration in both phases. Intuitively, since SWCNTs coated with SC prefer the bottom phase, one could expect that the SC molecules themselves would have a higher affinity to the bottom phase and that is where they should be more abundant.

Considering that the distribution of SWCNTs in biphasic systems is strongly dependent on the type of phase-forming polymers, their molecular weights, and concentrations[14,26], we performed additional measurements in PEG-DEX systems of different proportions of DEX and PEG. The goal was to gain further insight how the surfactant distribution changes upon modifying the extraction conditions. Regardless whether DEX (**Figure 2e** – B1) or PEG **(Figure 2e** – C1) was twice as concentrated as the other phase-forming component, the amount of SC in the top and bottom phases was analogous. The small discrepancy (experimental uncertainty, presented as shaded area in all the optical absorption spectra enclosed in this article) was not statistically significant (**Table S4**). Hence, in a PEG-DEX system containing a single type of surfactant, the obtained results strongly suggest that it is homogeneously partitioned between the two phases.

However, the ATPE systems achieve their best resolution for SWCNT separation when multiple surfactants are used to enable their competitive adsorption on SWCNT surface[6]. Since, as previously discussed, bile salt surfactants such as SC direct SWCNTs to the bottom, Triton X-100 (TX-100, a non-ionic surfactant)[10] and sodium dodecyl sulfate (SDS, an anionic surfacant)[27–29] were examined due to their capacity to act as counter-surfactants preferring SWCNT migration to the top. In both cases, despite the presence of these compounds in the extraction system, the concentration of SC was again not statistically different between the top and bottom phases **(Figure S5, Figure S6)**.

Furthermore, we also evaluated other partitioning systems to prove the utility of the newly developed method for in situ determination of SC concentration. For this purpose, we replaced either DEX or PEG in the DEX/PEG system with Ficoll or two types of Pluronic and quantified the amount of SC therein, respectively (**Figure S7**). In the case of the Ficoll/PEG system, the concentration of SC was analogous in both phases once more proving that under such conditions SC does not tend to favor any of the phases. However, the application of DEX/Pluronic systems (based on Pluronic P68 and Pluronic L35) led for the first time to the dissimilar concentration of SC between the top and bottom phases. The reduced amount of SC in the bottom phase finally explains why in DEX/Pluronic systems, the migration of SWCNTs to the bottom phase is more challenging than when classical PEG/DEX systems are employed[29]. This effect, probably originating from the possibility of the formation of mixed micelle systems between Pluronic and SC molecules in the top phase[30] is explained in detail in SI (**Figures S8-S11**).

To interpret the reaction between the bile salt surfactant SC and (4-OH)BA, which is at the heart of the reported method, we systematically examined a broad spectrum of bile salt derivatives (**Figure 3a**). While the core of these compounds is similar, they differ in the position and the number of hydroxyl groups attached as well as on the chemical identity of the joined chain constituting their hydrophilic head.

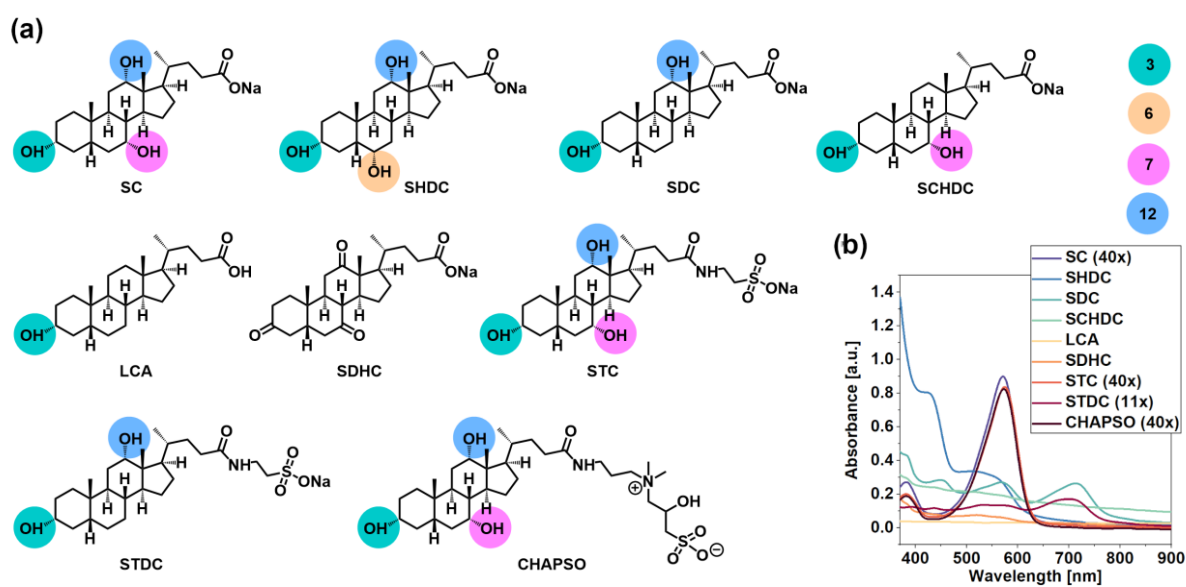

**Figure 3** (a) Bile salt derivatives examined, (b) optical absorption spectra of the products of reactions between the indicated compounds and (4-OH)BA (dilution factors in brackets).

Among them, only SC, STC, and CHAPSO produced single optical absorption features, the intensities of which were very high. To enable comparison between the bile salt derivatives, the samples had to be diluted 40 times. Interestingly, while the chemical structure of the tail did not matter, the presence of hydroxyl groups in 7- and 12-positions was crucial to produce the desired colored product enabling their optical detection. It is not possible to determine how important is the hydroxyl moiety in the 3-position because such compounds are simply not produced by nature.

Unfortunately, the elucidation of the reaction product is also highly problematic. We tried to isolate the reaction product, however, it is only stable in acidic conditions established by the reaction mixture. Attempts to dilute the mixture or transfer it to another solvent failed and resulted in product degradation, judging by discoloration of the solution and formation of a precipitate. Still, there are multiple chemical tests for the analysis of saccharides, which can reveal some insight into the underlying chemistry. It is known that aromatic aldehydes such as FUR formed during dehydration of sugars react easily with various hydroxyl-containing compounds in acidic environment to give characteristic colored products. An example is the Seliwanoff's test, wherein the aromatic aldehyde reacts with two resorcinol molecules, forming a coloured three-membered ring product[31]. Similarly, Bial's test based on orcinol[32] and Molisch's test based on napthol[33] also transform aromatic aldehydes into particular colored products. The best performing (4-OH)BA aromatic aldehyde is therefore capable of reacting with sodium cholate and its derivatives, both of which contain the necessary hydroxyl groups for the chemical transformation to proceed. In parallel, it is also highly probable that in an acidic environment the 3,7,12-trihydroxycholic framework can be dehydrated and a cation can be formed, similarly to the Liebermann–Burchard reaction, used to detect cholesterol[34]. This would explain why the product is stable only under acidic conditions and decomposes during attempts to dilute it with water or change the solvent.

Finally, capitalizing on the fresh insight about the behavior of surfactants in biphasic systems, it is worthwhile to improve the understanding of the SWCNT partitioning with this method. SWCNTs injected into a PEG-DEX system containing SDS are initially positioned in the top phase. The addition of SC, as discovered in this work, leads to equal concentration of SC in the top and bottom phases **(Figure 4)**. A gradual increase in SC concentration increases the coverage of SWCNTs with surfactant molecules, thereby decreasing the solvent accessible surface area. In turn, this promotes the migration of SWCNTs to the more hydrophilic bottom phase since the hydrophobic nature of SWCNTs is better concealed. Bile salt surfactants such as SC create relatively tight coatings around SWCNTs[35], which enhances this effect and increases the number of hydrophilic groups attached to SC, which are physically positioned on the SWCNT surface, facilitating their suspension in the more hydrophilic dextran-rich bottom phase. Besides, since the bottom phase has a higher density, the increased density of the closely wrapped SWCNTs with SC further improve the odds of their migration to the bottom phase purely due to density[36]. On the other hand, SDS creates loose coatings around SWCNTs[29,37] enabling their better contact with the solvent, which explains why SWCNTs encapsulated with SDS prefer the more hydrophobic top phase. Simultaneously, these results once again prove that the dynamic exchange of surfactant molecules occurs on the SWCNT surface and this phenomenon can be amplified by increasing the concentration of the desired surfactant type.

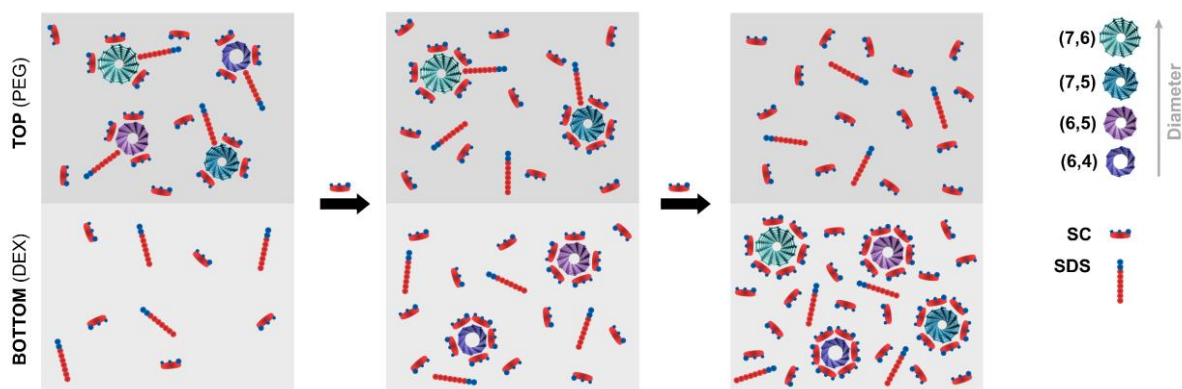

**Figure 4** The hypothesized partitioning mechanism of SWCNTs facilitated by bile salt surfactants.

**Conclusions**

Despite considerable progress, the mechanism of both the ATPE method discovered in 1896 and the Pettenkofer's test devised in 1844 remain elusive. However, the implementation and improvement of the latter approach carried out in this work enabled us to gain a substantial insight into the principles governing the biphasic extraction. Concomitantly, a straightforward test for the optical detection of bile salt surfactants, commonly used for the dispersion of otherwise insoluble nanomaterials, can be very useful to better understand their behavior in liquid media. The technique proved to work successfully in complex chemical media, enabling precise quantification of such compounds due to the linear dependence of optical absorption of the generated, highly-colorful products on surfactant concentration. In the context of SWCNTs, the obtained knowledge can facilitate the development of more effective methods of their purification, which is necessary to unleash their unique properties beyond the laboratory environment.

**Acknowledgments**

The authors would like to thank the Polish National Science Centre for the financial support of the research (under the OPUS program, Grant agreement 2019/33/B/ST5/00631).

Supplementary Information file

# Colorization of Optically Transparent Surfactants to Track Their Movement in Biphasic Systems Used for Differentiation of Nanomaterials

Błażej Podleśny[a,*], Łukasz Czapura[a], Dawid Janas[a,*]

[a] *Department of Organic Chemistry, Bioorganic Chemistry and Biotechnology, Silesian University of Technology, B. Krzywoustego 4, 44–100 Gliwice, Poland*

[*] Corresponding authors: Blazej.Podlesny@polsl.pl (B.P), Dawid.Janas@polsl.pl (D.J.)

**Table of Contents**

# 1. Experimental

## 1.1. Materials

**Table S1** Compounds that were used to prepare samples, their acronyms and vendors.

| Compound | Acronym/formula | Vendor |
|---|---|---|
| 2-bromobenzaldehyde | (2-Br)BA | Sigma Aldrich (USA) |
| 2-nitrobenzaldehyde | (2-$NO_2$)BA | Sigma Aldrich (USA) |
| 3-([3-Cholamidopropyl]dimethylammonio)-2-hydroxy-1-propanesulfonate | CHAPSO | Abcr (Germany) |
| 3-hydroxybenzaldehyde | (3-OH)BA | Acros Organics (India) |
| 3-methoxybenzaldehyde | (3-OMe)BA | Acros Organics (India) |
| 4-hydroxybenzaldehyde | (4-OH)BA | Acros Organics (India) |
| Acetic acid (99.9%) | $CH_3COOH$ | VWR Chemicals (France) |
| Dextran (70 000 Da) | DEX | PanReac AppliChem (Germany) |
| Ficoll 400 | Ficoll | PanReac AppliChem (Germany) |
| Furfural | FUR | Thermo Scientific (Dominican Republic) |
| Lithocholic acid | LCA | Thermo Fisher (Italy) |
| Pluronic F68 | F68 | Sigma Aldrich (USA) |
| Pluronic L35 | P35 | Sigma Aldrich (USA) |
| Polye(ethylene glycol) (6 000 Da) | PEG | Alfa Aesar (Germany) |
| Signis SG65i SWCNTs | SWCNTs | Sigma Aldrich (USA) |
| Sodium chenodeoxycholate | SCHDC | Abcr (Germany) |
| Sodium cholate | SC | Alfa Aesar (New Zealand) |
| Sodium dehydrocholate | SDH | Abcr (Germany) |
| Sodium deoxycholate | SDC | Thermo Scientific (New Zealand) |
| Sodium dodecyl sulfate | SDS | Sigma Aldrich (USA) |
| Sodium hyodeoxycholate | SHDC | Abcr (Germany) |
| Sodium taurocholate hydrate | STC | Abcr (Germany) |
| Sodium taurodeoxycholate | STDC | Abcr (Germany) |
| Sucrose | $C_{12}H_{22}O_{11}$ | Südzucker (Poland) |
| Sulfuric acid (98%) | $H_2SO_4$ | Supelco (Belgium) |
| Triton X-100 | TX | Acros Organics |

Double-distilled water obtained from the Elix Millipore system was used for all experiments.

## 1.2. Preparation of SWCNT dispersions

SWCNTs (40 mg) and SC solution (40 mL; 2%, m/V) were introduced into a 50 mL vial and placed in an ice bath. Then, the mixture was sonicated (Hielscher UP200St, 200 min, 30 W) and centrifuged (Eppendorf Centrifuge 5804 R) at 18 °C at the Relative Centrifugal Force (RCF) of 15314 × g for 1.5 h to precipitate bundled SWCNTs. 80% of the upper volume of the supernatant was collected and used for further experiments. Optical absorption spectrum and PL map of as-made SWCNT dispersion are shown below (**Figure S1**).

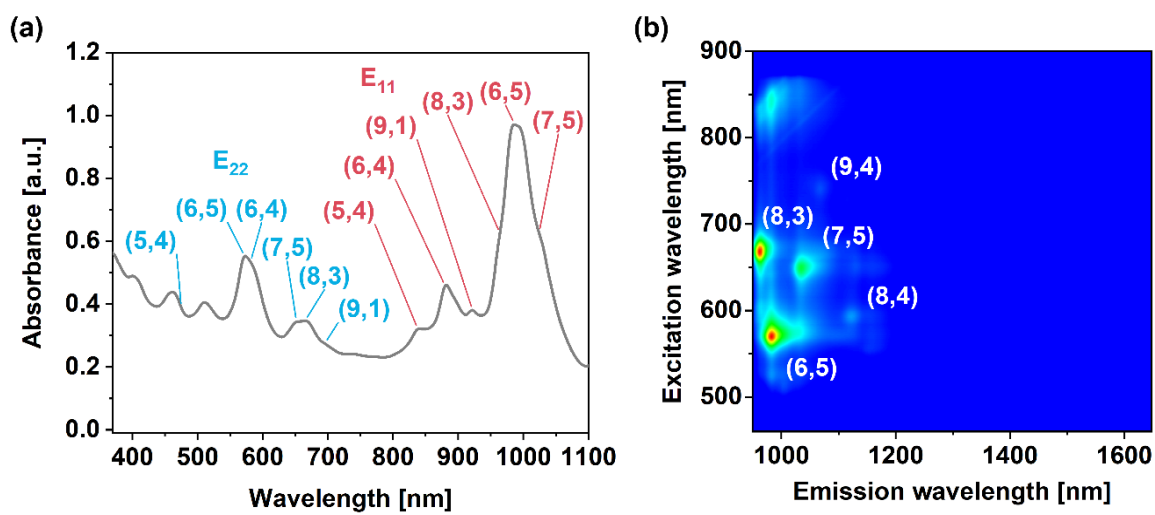

**Figure S1** Characterization of the starting SWCNT material dispersed in 2% SC: (a) Optical absorption spectrum indicating key chiralities (metallic species were not assigned) and (b) photoluminescence excitation-emission map (some of SWCNT types from the optical absorption spectrum were not detected due to limited measurement range or low photoluminescence quantum yield of selected SWCNT chiralities).

## 1.3. ATPE protocol

Water, stock solutions of phase-forming polymers (DEX/Ficoll and PEG/P68/P35/P121) and other components (surfactants and SWCNT dispersion) were transferred to a centrifuge tube. The mixture was then homogenized by a Vortex mixer (about 10 s per sample) and centrifuged (Eppendorf Centrifuge 5804 R) for 3 minutes at 18 °C at an RCF of 2025 × g to promote the formation of a biphasic system. Top and bottom phases were collected by pipetting to avoid cross-contamination.

Concentrations of stock solutions employed for ATPE sample preparation are included **in Table S2**. Compositions of ATPE samples are described in **Table S4** (for DEX-PEG system including only SC addition), **Table S7** (for DEX-PEG system including TX and SC addition), **Table S10** (for DEX-PEG system including SDS and SC addition) and **Table S13** (for other ATPE systems including SC addition)

**Table S2** Concentrations of stock solutions employed for ATPE sample preparation and determination of SC concentrations therein.

| Compound | DEX | PEG | SC | TX100 | SDS | Ficoll | P68 | P35 | P121 |
|---|---|---|---|---|---|---|---|---|---|
| Concentration of stock solutions (w/w; water) [%] | 20 | 50[1] | 10[2] | 2.5 | 10 | 20 | 25 | 25 | 25 |

[1] 25% where indicated
[2] 2% when used instead of SWCNT dispersion

### 1.4. Determination of binodal curve

In a clean and dry glass Nessler cylinder, a magnetic stirring bar was placed, after which the combined weight was recorded. Subsequently, a sufficient amount of DEX solution (70 kDa) was added, and the weight of the system was measured again. Afterward, PEG (6 kDa) was introduced dropwise while stirring until the solution became cloudy. The weight of the mixture and the glassware was re-registered. Next, water was slowly added until the solution was clear again, and the weight was acquired. This routine leading to a cloudy/transparent solution was repeated several times to collect more measurement points. Finally, based on the obtained data, the mass fractions of both polymers necessary to form the biphasic system was drawn in the form of a binodal curve (**Figure 2d**).

### 1.5. Optical characterization

Optical absorption spectra were measured with the Hitachi U2910 spectrophotometer were measured from 360 to 1100 nm using cuvettes made of polystyrene with the optical path length of 10 mm. A reference cuvette containing distilled water was put in the second measurement channel for analysis. Excitation-emission photoluminescence (PL) maps were measured with PhotonClaIR spectrophotometer across the specified wavelengths ranges: excitation (460–900 nm) and emission (948–1600 nm).

### 2. Determination of SC concentration in aqueous solution

### 2.1. The reaction procedure developed by Pettenkofer

In 1844, Pettenkofer reported that mixing sugar, concentrated sulfuric acid and ox bile resulted in the appearance of a violet color, reminiscent of potassium permanganate. Based on this observation, he proposed that this method could be used for detecting cholic acid. The procedure involves careful addition of concentrated sulfuric acid (approximately two-thirds the volume of the bile acid solution) to a small portion of the analyte, ensuring the temperature does not exceed 62.5°C. Subsequently, two or three drops of a cane sugar solution (prepared by dissolving one part sugar in four to five parts water) are added. The mixture is then shaken gently, and if the cholic acid is present, purple color appears[1]. We adopted this strategy to evaluate whether Pettenkofer's procedure could be used to detect sodium salt of cholic acid dissolved in water, which is commonly used to improve the partitioning of materials within the framework of the ATPE method. After combining 2 mL each of aqueous solutions of SC (2%) and sucrose (20%), 4 mL of concentrated sulfuric acid was slowly introduced down the wall of the glass cylinder. Due to the density difference, two phases were formed, and subsequently yellowish color started to appear at the interface **(Figure 2)**. After a few minutes, the color at the interface turned purple indicating that Pettenkofer's test can be utilized to determine SC presence. In a reference experiment, wherein SC solution was replaced with water, the interface became black because of the formation of the products of sucrose decomposition.

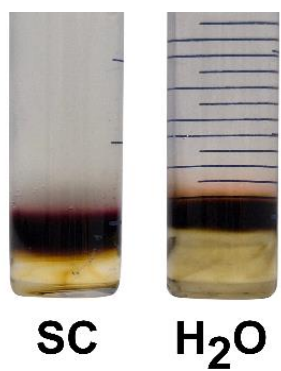

**Figure S2** Picture of samples subjected to Pettenkofer's test in the presence/absence of SC.

## 2.2. Improved protocol reported in this work

According to the results described in the main text, the application of an aromatic aldehyde in the form of (4-OH)BA instead of sucrose produced the best results, so this reactant was used for the quantification of SC. In such a case, 200 µL of 1% (m/m) solution of (4-OH)BA in 70% acetic acid was mixed with 200 µL of 70% acetic acid and 100 µL of the analyzed cholate-containing sample in a glass container. The vial was then placed in a glycerin bath preheated to 70°C. Once the reagent mixture reached 70°C, 2.6 mL of sulfuric acid (1:1.3 v/v corresponding to 57% m/m) was added. The mixture was subsequently incubated at this temperature for 10 minutes, after which it was transferred to an ice bath to halt the reaction. After cooling, the mixture was stored in the dark to allow the reaction to develop, which led to more intense coloration of the sample, giving to higher absorbance values, thereby facilitating cholate detection. We noted that the 7 days of incubation time was sufficient.

To validate this approach, we carried out a series of six experiment series, each of which examined a number of SC concentration values under various ambient conditions. All of the calibration curves shown below had very good linearity with high goodness of fit, which approached unity **(Figure S3)**. Besides, all of them were practically analogous and no major deviation from the trend could be observed. As a result, the proposed approach may be used for quantitative analysis with confidence.

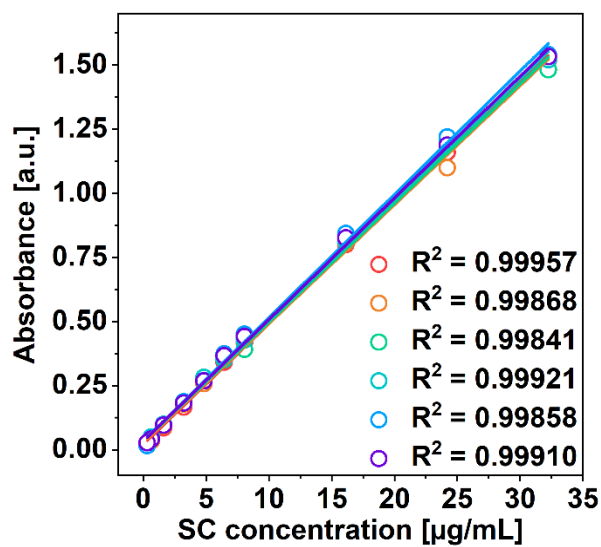

**Figure S3** Intensity of optical absorption at the maximum of the peak located at ca. 567 nm as a function of SC concentration after 7-day reaction time.

3.    Determination of SC concentration in DEX-PEG systems

**Table S3** Compositions of ATPE samples prepared using the DEX-PEG system and SC.

| Sample | DEX$_{aq.}$ (20%) [µL] | PEG$_{aq.}$ (50%) [µL] | SC$_{aq.}$ (10%) [µL] | SC$_{aq.}$ (2%) [µL] | 1 mg/mL SWCNT dispersion in SC$_{aq.}$ (2%) [µL] | H$_2$O [µL] |
|---|---|---|---|---|---|---|
| A1 | 1350 | 540 | - | - | - | 2700 |
| A2 | 1350 | 540 | 360 | - | - | 2340 |
| A3 | 1350 | 540 | 360 | 225 | - | 2115 |
| A4 | 1350 | 540 | 360 | - | 225 | 2115 |
| B1 | 1380 | 1110 | 360 | - | - | 1740 |
| B2 | 1380 | 1110 | 360 | 225 | - | 1515 |
| C1 | 2760 | 550 | 360 | - | - | 920 |
| C2 | 2760 | 550 | 360 | 225 | - | 695 |

The spectra registered for these samples, after subjecting them to the modified Pettenkofer procedure, can be found in **Figure 2a** and **Figure S4**. Based on the acquired spectral data and the calibration curve (**Figure S3**), the mass of SC in each phase was calculated (**Table S4**).

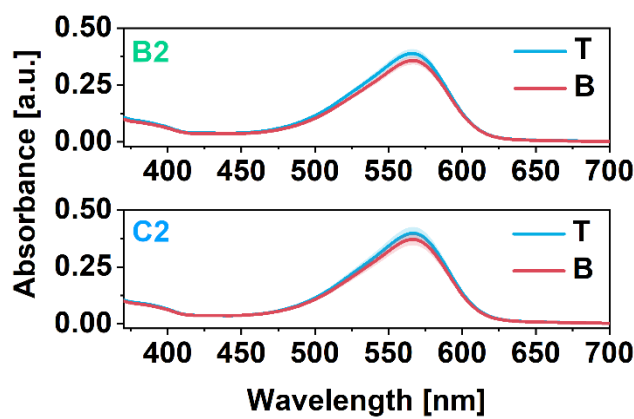

**Figure S4** Optical absorption spectra of top and bottom phases of the samples prepared in the indicated conditions (**Table S3**) after subjecting them to the modified Pettenkofer procedure.

**Table S4** Determined mass of SC in the top and bottom phases for the indicated samples using the calibration curve displayed in **Figure S3**. The discrepancy between the calculated and introduced mass of SC was used for the presentation of statistical uncertainty as shaded areas in **Figure 2e** and **Figure S4**.

| Sample | Mass of SC | | | | |
|---|---|---|---|---|---|
| | Mass of SC introduced into the sample [mg] | Calculated mass of SC in top phase | Calculated mass of SC in bottom phase | Calculated mass of SC in both phases | Calculated mass/ introduced mass [%] |
| A1 | - | - | - | - | - |
| A2 | 36.00 | 25.64 | 11.52 | 37.16 | + 3% |
| A3 | 40.50 | 29.97 | 13.10 | 43.08 | + 6% |
| A4 | 40.50 | 30.28 | 13.40 | 43.68 | + 8% |
| B1 | 36.00 | 25.28 | 11.00 | 36.29 | + 1% |
| B2 | 40.50 | 29.86 | 12.18 | 42.04 | + 5% |
| C1 | 36.00 | 27.19 | 11.03 | 38.22 | + 6% |
| C2 | 40.50 | 30.66 | 12.69 | 43.35 | + 7% |

## 4. Determination of SC concentration in DEX-PEG systems with TX-100

Analogous experiments were conducted in the DEX-PEG system operated with the addition of TX-100 as a counter-surfactant favoring the migration of SWCNTs to the top phase. The parameters used to carry out purification are provided in **Table S5**, whereas the corresponding optical absorption spectra and the mass of SC determined in the relevant phases is given in **Figure S5** and **Table S6**, respectively.

**Table S5** Compositions of ATPE samples prepared using the DEX-PEG system as well as SC and TX-100.

| Sample | DEX$_{aq.}$ (20%) [μL] | PEG$_{aq.}$ (50%) [μL] | SC$_{aq.}$ (10%) [μL] | SC$_{aq.}$ (2%) [μL] | TX-100$_{aq.}$ (2.5%) [μL] | 1 mg/mL SWCNT dispersion in SC$_{aq.}$ (2%) [μL] | H$_2$O [μL] |
|---|---|---|---|---|---|---|---|
| D1 | 1350 | 540 | - | - | 650 | - | 2050 |
| D2 | 1350 | 540 | 360 | - | 650 | - | 1690 |
| D3 | 1350 | 540 | 360 | 225 | 650 | - | 1465 |
| D4 | 1350 | 540 | 360 | - | 650 | 225 | 1465 |

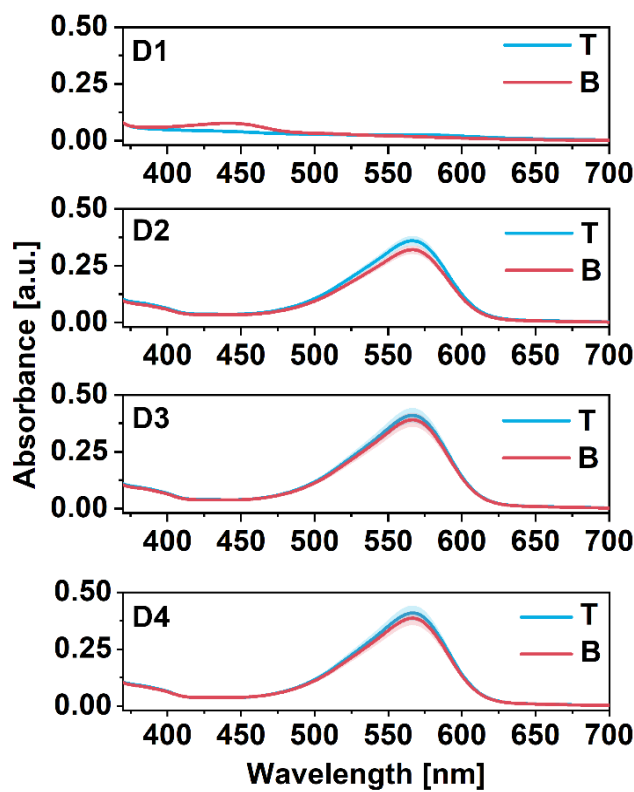

**Figure S5** Optical absorption spectra of top and bottom phases of the samples prepared in the indicated conditions (**Table S5**) after subjecting them to the modified Pettenkofer procedure.

**Table S6** Determined mass of SC in the top and bottom phases for the indicated samples using the calibration curve displayed in **Figure S3**. The discrepancy between the calculated and introduced mass of SC was used for the presentation of statistical uncertainty as shaded areas in **Figure S5**.

| Sample | Mass of SC | | | | |
|---|---|---|---|---|---|
| | Mass of SC introduced into the sample [mg] | Calculated mass of SC in top phase | Calculated mass of SC in bottom phase | Calculated mass of SC in both phases | Calculated mass/ introduced mass [%] |
| D1 | - | - | - | - | - |
| D2 | 36.00 | 27.23 | 10.79 | 38.06 | + 6% |
| D3 | 40.50 | 30.76 | 13.03 | 43.79 | + 8% |
| D4 | 40.50 | 30.71 | 12.87 | 43.58 | + 8% |

5. **Determination of SC concentration in DEX-PEG systems with SDS**

Similar experiments were conducted in the DEX-PEG system operated with the addition of SDS as a counter-surfactant favoring the migration of SWCNTs to the top phase. The parameters used to carry out purification are provided in **Table S7**, whereas the corresponding optical absorption spectra and the mass of SC determined in the relevant phases is given in **Figure S6** and **Table S8**, respectively.

**Table S7** Compositions of ATPE samples prepared using the DEX-PEG system as well as SC and SDS.

| Sample | DEX$_{aq.}$ (20%) [μL] | PEG$_{aq.}$ (50%) [μL] | SC$_{aq.}$ (10%) [μL] | SDS$_{aq.}$ (10%) [μL] | 1 mg/mL SWCNT dispersion in SC$_{aq.}$ (2%) [μL] | H$_2$O [μL] |
|---|---|---|---|---|---|---|
| E1 | 1350 | 540 | - | 180 | - | 2520 |
| E2 | 1350 | 540 | 360 | 180 | - | 2160 |
| E3 | 1350 | 540 | 360 | 180 | 225 | 1935 |

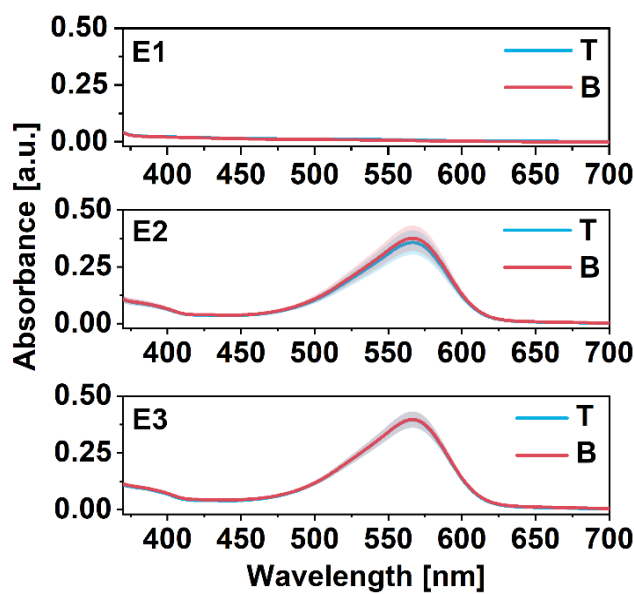

**Figure S6** Optical absorption spectra of top and bottom phases of the samples prepared in the indicated conditions (**Table S7**) after subjecting them to the modified Pettenkofer procedure.

**Table S8** Determined mass of SC in the top and bottom phases for the indicated samples using the calibration curve displayed in **Figure S3**. The discrepancy between the calculated and introduced mass of SC was used for the presentation of statistical uncertainty as shaded areas in **Figure S6**.

| Sample | Mass of SC | | | | |
|---|---|---|---|---|---|
| | Mass of SC introduced into the sample [mg] | Calculated mass of SC in top phase | Calculated mass of SC in bottom phase | Calculated mass of SC in both phases | Calculated mass/ introduced mass [%] |
| E1 | - | - | - | - | - |
| E2 | 36.00 | 28.03 | 13.20 | 41.23 | + 15% |
| E3 | 40.50 | 30.51 | 13.66 | 44.17 | + 9% |

6.     Determination of SC concentration in other partitioning system

Analogous experiments were conducted in other partitioning systems wherein either PEG or DEX was substituted for another phase-forming polymer. The parameters used to carry out purification are provided in **Table S9**, whereas the corresponding optical absorption spectra and the mass of SC determined in the relevant phases is given in **Figure S7** and **Table S10**, respectively.

Table S9 Compositions of other system ATPE samples

| Sample | DEX (20%) [μL] | PEG (25%) [μL] | Ficoll (20%) [μL] | PL68 (20%) [μL] | PL35 (20%) [μL] | SC (10%) [μL] | H$_2$O [μL] |
|---|---|---|---|---|---|---|---|
| F1 | - | 2000 | 2000 | - | - | 360 | 230 |
| F2 | 2000 | - | - | 2000 | - | 360 | 230 |
| F3 | 2000 | - | - | - | 2000 | 360 | 230 |

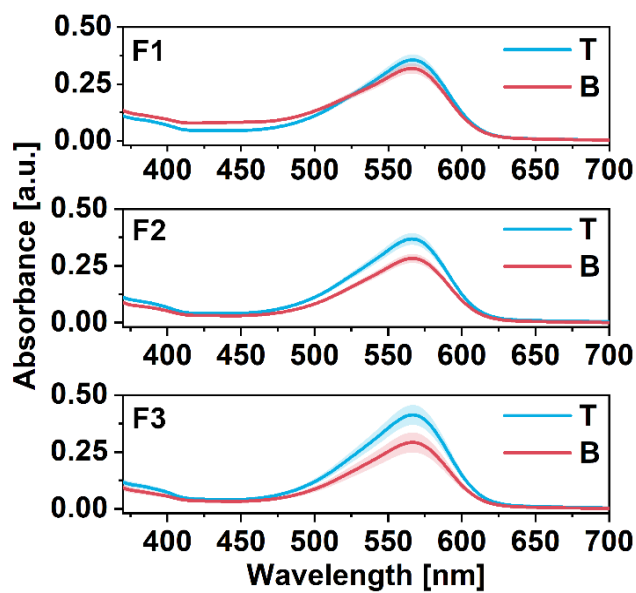

**Figure S7** Optical absorption spectra of top and bottom phases of the samples prepared in the indicated conditions (**Table S9**) after subjecting them to the modified Pettenkofer procedure.

**Table S3** Determined mass of SC in the top and bottom phases for the indicated samples using the calibration curve displayed in **Figure S3**. The discrepancy between the calculated and introduced mass of SC was used for the presentation of statistical uncertainty as shaded areas in **Figure S7**.

| Sample | Mass of SC introduced into the sample [mg] | Mass of SC | | | |
|---|---|---|---|---|---|
| | | Calculated mass of SC in top phase | Calculated mass of SC in bottom phase | Calculated mass of SC in both phases | Calculated mass/ introduced mass [%] |
| F1 | 36.00 | 28.97 | 9.71 | 38.68 | + 7% |
| F2 | 36.00 | 28.97 | 9.71 | 38.68 | + 7% |
| F3 | 36.00 | 31.61 | 9.68 | 41.29 | + 15% |

## 7. Comparison of DEX-PEG and DEX-PL partitioning systems

Four extraction systems were prepared to investigate the difference in the capacity of SC to cause downward migration of SWCNTs. The optical absorption spectra along with photographs of the analyzed samples are provided in **Figures S8-S11**. The parameters used for the separations are given in captions accompanying the graphics.

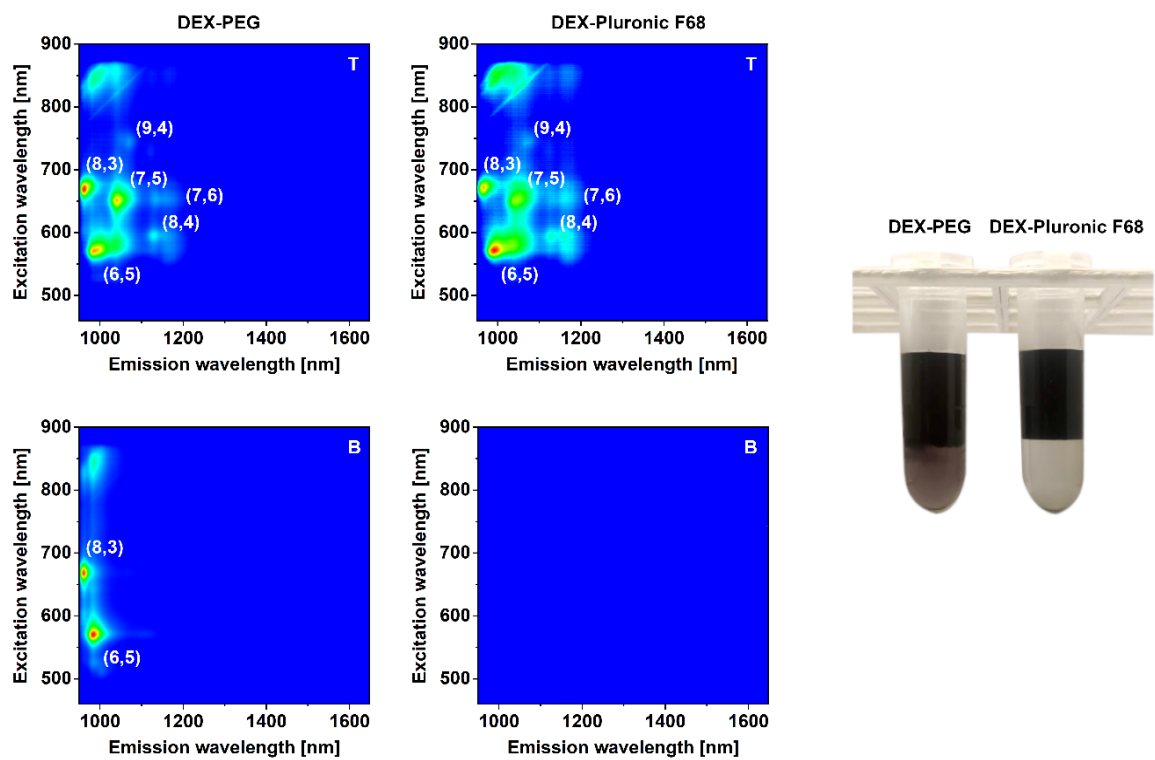

**Figure S8** Photoluminescence excitation-emission maps and photographs of SWCNT samples sorted in DEX-PEG and DEX-Pluronic F68 systems. Extraction parameters: 360 µL of PEG or Pluronic F68 (25%), 450 µL of DEX (24%), 75 µL of SWCNT dispersion (in 2% SC solution), and 645 µL of $H_2O$.

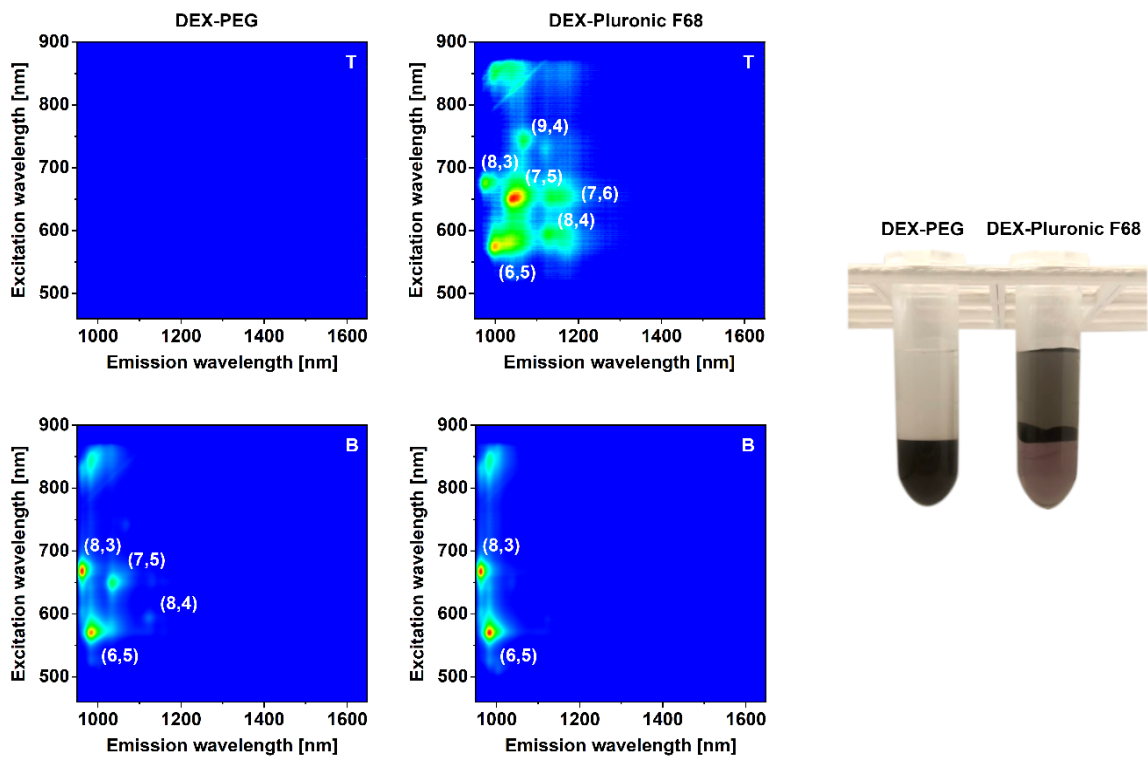

**Figure S9** Photoluminescence excitation-emission maps and photographs of SWCNT samples sorted in DEX-PEG and DEX-Pluronic F68 systems. Extraction parameters: 360 μL of PEG or Pluronic F68 (25%), 450 μL of DEX (24%), 75 μL of SWCNT dispersion (in 2% SC solution), 120 μL of SC (10%), and 525 μL of $H_2O$.

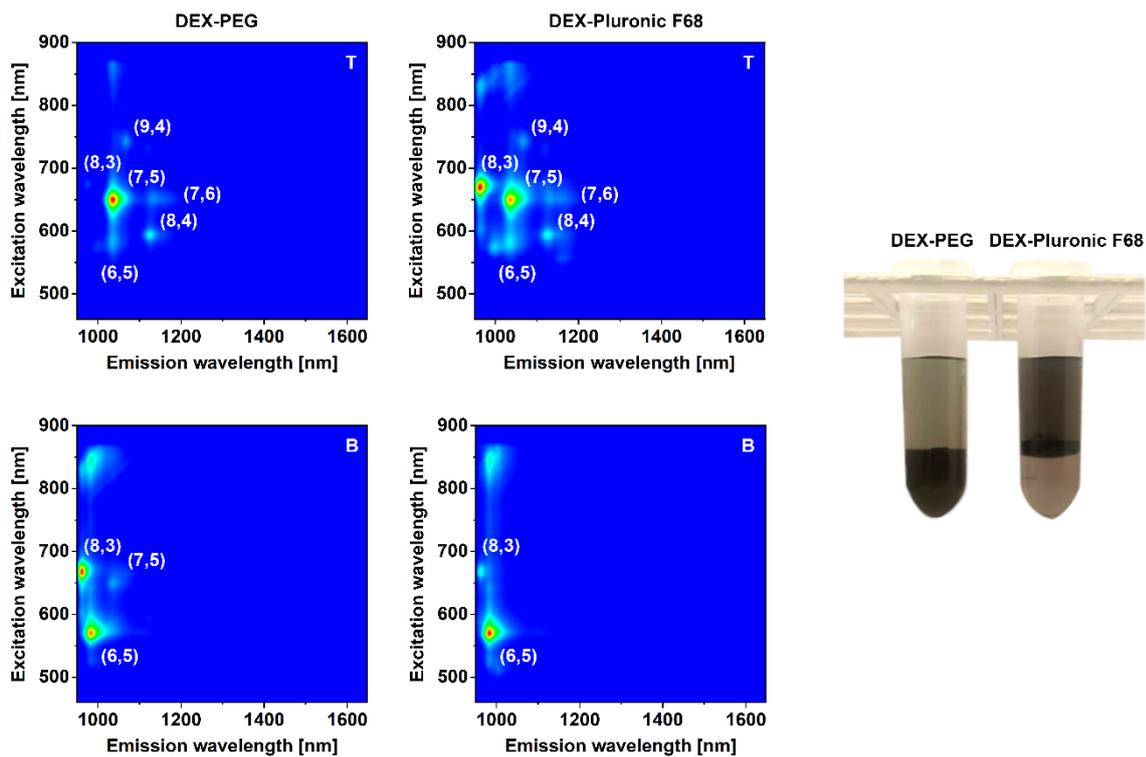

**Figure S10** Photoluminescence excitation-emission maps and photographs of SWCNT samples sorted in DEX-PEG and DEX-Pluronic F68 systems. Extraction parameters: 360 µL of PEG or Pluronic F68 (25%), 450 µL of DEX (24%), 75 µL of SWCNT dispersion (in 2% SC solution), 120 µL of SC (10%), 35 µL of TX-100 (2.5%), and 490 µL of $H_2O$.

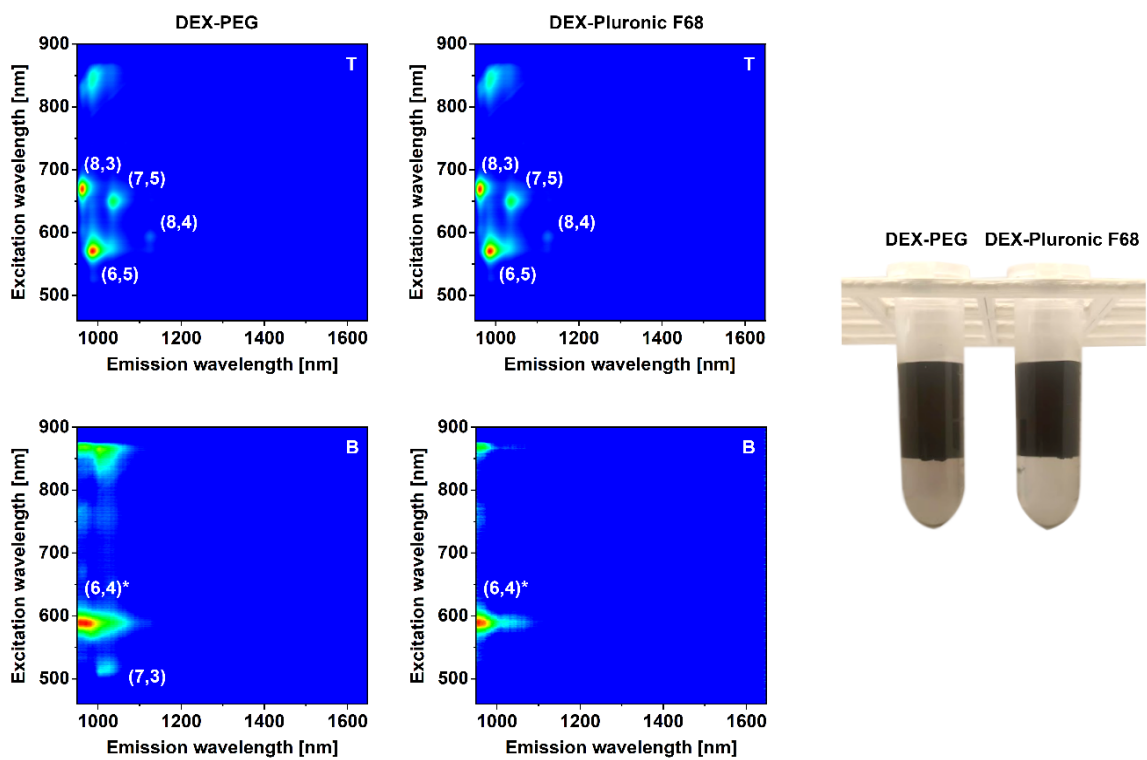

**Figure S11** Photoluminescence excitation-emission maps and photographs of SWCNT samples sorted in DEX-PEG and DEX-Pluronic F68 systems. Extraction parameters: 360 μL of PEG or Pluronic F68 (25%), 450 μL of DEX (24%), 75 μL of SWCNT dispersion (in 2% SC solution), 120 μL of SC (10%), 200 μL of TX-100 (2.5%), and 325 μL of H$_2$O. Due to the limitations of the instrument, only the defect-induced $E_{11}^*$ peak in (6,4) SWCNTs is observed[2].

The presented results clearly show that the same amount of SC added into the biphasic system promotes the migration of SWCNTs to the bottom phase to a lower extent in the case of DEX-Pluronic partitioning systems. While we cannot eliminate from consideration the surface-active properties of Pluronic molecules, which may affect the SWCNT differentiation course, the lower amount of SC in the bottom phase, as shown and quantified in **Figure S7** and **Table S10** (**Samples F2 and F3**), respectively, can also hinder the downward extraction of SWCNTs.